\documentclass[12pt]{iopart}

\usepackage{iopams}
\usepackage{equations}
\usepackage{latexsym}
\usepackage{graphicx}
\usepackage[dvips]{color}
\usepackage{cite}
\renewcommand{\eref}[1]{Eq.~(\ref{#1})}
\renewcommand{\fref}[1]{Fig.~\ref{#1}}
\def\dfrac#1#2{{\displaystyle{#1 \over #2}}}
\def\dint{\displaystyle\int}

\def\e    {\mathrm{e}}

\def\pte  {p_{te}}
\def\ute  {u_{te}}
\def\vte  {v_{te}}
\def\Zeff {Z_\mathrm{eff}}
\def\Bmax {B_\mathrm{max}}
\def\Bmin {B_\mathrm{min}}

\def\bVdr {{\bf V}_{\!\mathrm{dr}}}
\def\Vdr  {V_{\!\mathrm{dr}}}
\def\vpar {v_\parallel}

\def\upar {u_\parallel}

\def\cL   {\mathcal{L}}
\def\cV   {\mathcal{V}}

\def\cR   {\mathcal{R}}
\def\cO   {\mathcal{O}}

\begin{document}

\title[Relativistic neoclassical radial fluxes]
{Relativistic neoclassical radial fluxes in the $1/\nu$ regime}

\author{I.~Marushchenko$^1$, N.~A.~Azarenkov$^1$, N.~B.~Marushchenko$^2$}

\address{$^1$ V.~N.~Karazin Kharkiv National University, Svobody Sq.~6, 61022, 
Kharkiv, Ukraine}
\address{$^2$ Max Planck Institute for Plasma Physics, EURATOM Association,
         Wendelsteinstr.~1, 17491 Greifswald, Germany}
\ead{i.marushchenko@gmail.com}

\begin{abstract}
The radial neoclassical fluxes of electrons in the $1/\nu-$regime are 
calculated with relativistic effects taken into account and compared 
with those in the non-relativistic approach. 
The treatment is based on the relativistic drift-kinetic equation with the 
thermodynamic equilibrium given by the relativistic Maxwell-J\"uttner 
distribution function. 
It is found that for the range of fusion temperatures, $T_e<100$~keV,   
the relativistic effects produce a reduction of the radial fluxes which does 
not exceed 10\%.
This rather small effect is a consequence of the non-monotonic temperature 
dependence of the relativistic correction caused by two counteracting factors: 
a reduction of the contribution from the bulk and a significant broadening 
with the temperature growth of the energy range of electrons contributing 
to transport.

The relativistic formulation for the radial fluxes given in this paper is 
expressed in terms of a set of relativistic thermodynamic forces which is 
not identical to the canonical set since it contains an additional relativistic 
correction term dependent on the temperature.
At the same time, this formulation allows application of the non-relativistic 
solvers currently used for calculation of mono-energetic transport coefficients.
\end{abstract}

\submitto{\PPCF}
\maketitle

\section{Introduction}
\label{sec:Intr}

The role of relativistic effects in hot plasmas has been recognized as 
important not only in astrophysics 
\cite{deGroot:RelativKineticTheory_1980,TenBargeHazeltineMahajan:PoP2008}
but also in fusion, in particular, for the population of highly energetic 
runaway electrons in tokamaks \cite{ConnorHastie:NF1975}.
However, the relativistic effects do not necessarily require an extremely 
high temperatures since they can be non-negligible even if $T_e$ is only 
on the order of tens of keV, i.e. $T_e \ll m_{e0}c^2$. 
These effects appear due to the macroscopic features of the relativistic 
thermodynamic equilibrium given by the Maxwell-J{\"u}ttner distribution 
function \cite{deGroot:RelativKineticTheory_1980,BraamsKarney:PoFB1989}.
An example of such effects provided by the Maxwell-J{\"u}ttner distribution function 
is given in a recent paper \cite{Marushchenko_Pei:PoP2012}, where the stability 
criterion for collisional heat transfer from hot electrons to ions with respect 
to the Coulomb decoupling is studied and it is found that relativistic effects lead 
to qualitative changes in stability criteria.
While in non-relativistic plasmas criterion is given by $T_e/T_i<3$, relativistic 
effects makes it temperature-dependent and for $T_{e,i}>75$~keV 
the collisional coupling between electrons and ions becomes absolutely stable.

Relativistic effects in fusion are surely not important for the ions,
but the transport physics for electrons needs to be examined carefully 
for fusion reactor projects such as ITER 
\cite{Tomabechi_ITER:NF1991,Wagner:PPCF2010,Giruzzi:PPCF2011} and DEMO 
\cite{Ward:PPCF2010,Horton:FST2008}, in which the expected electron temperature 
is sufficiently high, $T_e\simeq$~20~--~50~keV, and for future aneutronic 
fusion reactors with D--$^3$He and may be $p\,$--$^{11}$B reactions, which require 
temperatures of up to 70~--~100~keV 
\cite{MantsinenSalomaa:FT1998,SonFisch:PLA2004,Stott:PPCF2005}. 
However, all transport codes (see, for example \cite{ASTRA:IPP5-98}) developed 
to date and applied for simulations of reactor scenarios are based on the 
non-relativistic approach. 
Furthermore, there is no quantitative definition of an applicability range 
for the non-relativistic transport models so far.

Relativistic kinetics and MHD in plasmas are usually treated in the covariant 
formulation \cite{deGroot:RelativKineticTheory_1980,TenBargeHazeltineMahajan:PoP2008}.
For neoclassical transport, however, the covariant formulation is not necessary 
since Lorentz invariance is of minor importance with respect to the characteristic 
drift velocity, $\Vdr/c\ll 1$.
For this purpose, one can directly apply the relativistic drift-kinetic 
equation \cite{Littlejohn:PoF1984} with the relativistic Coulomb operator 
\cite{BraamsKarney:PoFB1989}.

In this paper, the relativistic effects in the radial fluxes in the $1/\nu$-regime,
which might be the most dangerous regime for future burning plasmas in stellarators 
and where the radial electric field plays no significant role, 
are estimated.
This case was chosen for investigation because the role of the highly energetic 
tail of the distribution function in transport processes in this regime is 
expected to be the largest in comparison with other regimes.
Indeed, the diffusion coefficient in the $1/\nu$-regime scales roughly as 
$\Vdr^2/\nu_e\propto v^7$, while in the tokamak banana-regime it scales 
as $\rho_{ce}^2\nu_e\propto v^{-1}$ (here, $\Vdr$ is the radial drift-velocity, 
$\rho_{ce}$ is the Larmor radius and $\nu_e$ is the collision frequency).

In Sec.~\ref{sec:rDKE}, the relativistic drift-kinetic equation (rDKE) in 
the mono-energetic approach with a set of thermodynamic forces 
which differs from the canonical one is formulated.
Only radial gradients are taken into account while the parallel electric 
field is excluded from consideration.
In Sec.~\ref{sec:1nu}, rigorous expressions for the radial electron fluxes 
and transport coefficients in the $1/\nu-$regime are derived.
In particular, the expression for the relativistic radial heat flux is obtained.
As a guideline, the paper \cite{NemovKasilov:PoP1999} was used, where 
the same was calculated in the non-relativistic approach.
In Sec.~\ref{sec:compare}, the numerical comparison of the relativistic and 
non-relativistic transport coefficients and radial fluxes is performed, 
and in Sec.~\ref{sec:sum} a brief discussion of the results is given.

\section{Mono-energetic drift-kinetic equation for relativistic electrons}
\label{sec:rDKE}

The electron radial fluxes in toroidal plasmas (except the Ware pinch) can 
be calculated from the relativistic drift-kinetic equation (rDKE) for the 
first-order distribution function $f_{e1}$ in the mono-energetic approach 
\cite{HintonHazeltine:RMP1976, HelanderSigmar:CollisTransport_2002,DKES-1:PoF1986}. 
Using on a magnetic surface, with flux-sufrace label $\rho$, 
the set of variables $(s,u,\lambda)$, 
where $s$ is the coordinate along the field-line, $u=p/m_{e0}=\gamma v$ is 
the momentum per unit mass, $\gamma=\sqrt{1+u^2/c^2}$ is the Lorentz-factor, 
$\lambda=(1-\xi^2)/b$ is the normalized magnetic moment, where $\xi=\upar/u$ 
is the pitch and $b=B/B_0$ is the normalized magnetic field with the reference 
field $B_0$, the mono-energetic rDKE can be written as
\begin{equation}\label{eq:DKE1}
\cV(f_{e1})-\nu_D(u)\cL(f_{e1})=
-\left(\bVdr\cdot\nabla\rho\right)\,\dfrac{\partial F_{eMJ}}{\partial\rho}.
\end{equation}
The first term in \eref{eq:DKE1} is the mono-energetic Vlasov operator, 
$\cV=(\vpar{\bf h}+\bVdr)\cdot\nabla_s$, where ${\bf h}={\bf B}/B$ and $\nabla_s$ 
is the gradient within the magnetic surface (here, $\dot{\lambda}=0$).
The second term is the pitch-angle scattering operator with 
the deflection frequency $\nu_D(u)=\nu_D^{ee}(u)+\nu_D^{ei}(u)$ 
(the complete expressions for relativistic $\nu_D^{ee}$ and $\nu_D^{ei}$ are given 
in \ref{sec:nuDab}) and the Lorentz operator is
\begin{equation}\label{eq:Lorentz}
\cL=\dfrac{2\xi}{b}\dfrac{\partial}{\partial\lambda}
\left(\lambda \xi\dfrac{\partial}{\partial\lambda}\right).
\end{equation}
The relativistic drift velocity can be written as
\begin{equation}\label{eq:Vdr}
\bVdr=\dfrac{c}{B^2}\,{\bf E}\times{\bf B}-
\dfrac{m_{e0}cu^2(1+\xi^2)}{2e\gamma B^3}\,{\bf B}\times\nabla B
\end{equation}
with ${\bf E}=-\nabla\Phi=-\Phi^\prime\nabla\rho$ and 
$\Phi^\prime\equiv d\Phi/d\rho$, where $\Phi$ is the plasma potential 
(here and below, $e=|e|$).
One can see that only the last term in \eref{eq:Vdr} contributes to 
$\dot{\rho}\equiv\bVdr\cdot\nabla\rho$ on the right-hand side (RHS) 
of \eref{eq:DKE1}.
Since our treatment is limited to the $1/\nu$-regime, only such values 
of $E$ for which electrons with large $E/vB$ make no significant
contribution to transport are considered.			
In this case, the ${\bf E}\times{\bf B}$ drift term can be omitted 
in the Vlasov operator, i.e $\cV\simeq \vpar{\bf h}\cdot\nabla_s$.
(In the more general case, this term must be included to obtain the 
$\sqrt{\nu}$-regime which is more complex for analytical treatment 
and is not considered here.)

Thermodynamic equilibrium for relativistic electrons is given by the 
J\"uttner distribution function \cite{deGroot:RelativKineticTheory_1980} 
also known as the relativistic Maxwellian
\cite{BraamsKarney:PoFB1989}, which may be conveniently represented as
\begin{equation}\label{eq:fmxw}
f_{eMJ}(u,\rho)=\dfrac{n_e}{\pi^{3/2}\ute^3}C_{MJ}(\mu_r)
                \e^{-\mu_r(\gamma-1)},
\end{equation}
where $\ute=\pte/m_{e0}$ is the thermal momentum per unit mass with 
$\pte=\sqrt{2m_{e0}T_e}$ and $\mu_r=m_{e0}c^2/T_e$.
The Maxwell-J\"uttner distribution function is normalized by density, 
$n_e=\int d^3u\,f_{eMJ}$, and the normalization factor is
\begin{equation}\label{eq:CMJ}
C_{MJ}(\mu_r)=\sqrt{\dfrac{\pi}{2\mu_r}}
\dfrac{\e^{-\mu_r}}{K_2(\mu_r)}\simeq 1-\dfrac{15}{8\mu_r}+\cO(1/\mu_r^2),
\qquad\mu_r\gg 1,
\end{equation}
where $K_n(x)$ is the modified Bessel function of $n$-th order.
For convenience, the Maxwell-J\"uttner distribution function is used
in \eref{eq:DKE1} with the Boltzmann-factor included: 
\begin{equation}\label{eq:Fmxw}
F_{eMJ}=\e^{-e\Phi/T_e}f_{eMJ}.
\end{equation}

Since plasma parameters such as density and temperature 
only depend on the flux-surface label, $\rho$, the derivative in the right-hand-side 
of \eref{eq:DKE1} can be expressed in terms of the thermodynamic forces,
\begin{equation}\label{eq:RHS}
\dfrac{\partial F_{eMJ}}{\partial\rho}=
\left[A_1(\rho) + \kappa\,A_2(\rho)\right]F_{eMJ},
\end{equation}
where $\kappa=\mu_r(\gamma-1)$ is the relativistic kinetic energy normalized 
by $T_e$, and the thermodynamic forces $A_1$ and $A_2$ are defined as
\begin{subequations}
\label{eq:forcesA1A2}
\begin{eqnarray}
A_1(\rho) &=& \dfrac{n_e^\prime}{n_e}
         -\left(\dfrac{3}{2}+\cR\right)\dfrac{T_e^\prime}{T_e}
         -\dfrac{e\Phi^\prime}{T_e},     \label{forceA1} \\
A_2(\rho) &=& \dfrac{T_e^\prime}{T_e},   \label{forceA2}
\end{eqnarray}
\end{subequations}
with $n_e^\prime\equiv dn_e/d\rho$, $T_e^\prime\equiv dT_e/d\rho$, and 
the relativistic correction-term
\begin{equation}\label{eq:cR}
\cR(\mu_r)=\mu_r\left(\dfrac{K_3}{K_2}-1\right)-\dfrac{5}{2}
\simeq\dfrac{15}{8\mu_r}+\cO(1/\mu_r^2),\qquad\mu_r\gg 1.
\end{equation}
Note that in contrast to the ``canonical'' set of the thermodynamic forces 
\cite{HelanderSigmar:CollisTransport_2002,HintonHazeltine:RMP1976,DKES-1:PoF1986},
which depend only on the normalized gradients of density and temperature 
($n_e^\prime/n_e$ and $T_e^\prime/T_e$, respectively), and not on the absolute 
values of these plasma parameters, the first thermodynamic force $A_1(\rho)$ 
in the relativistic set \eref{eq:forcesA1A2} contains an additional 
temperature-dependent term.

Finally, the reduced mono-energetic rDKE can be represented as follows:
\begin{equation}\label{eq:DKEmono}
({\bf h}\cdot\nabla_s) f_{e1} - 
\dfrac{\gamma\nu_D(u)}{u\xi}\,\cL(f_{e1}) = 
-\dfrac{\gamma}{u\xi}\,\dot{\rho}\,
\left[A_1(\rho) + \kappa\,A_2(\rho)\right]F_{eMJ}.
\end{equation}

Note that similar to the non-relativistic formulation, the energy enters in 
\eref{eq:DKEmono} only as a parameter in $\gamma\nu_D(u)/u$ and the solution 
of \eref{eq:DKEmono} describes only the pitch- and spatial behavior of the 
distribution function $f_{e1}$, which is the same for both relativistic and 
non-relativistic approaches.
With the proper choice of parameters and right-hand-side of \eref{eq:DKEmono},
the solution from such solvers as DKES \cite{DKES-1:PoF1986} and NEO-2
\cite{Kernbichler_NEO-2:JJPFR2008}, which solve the non-relativistic DKE directly,
can be interpreted as a solution of the mono-energetic relativistic DKE.

\section{Relativistic radial fluxes}
\label{sec:1nu}

In this chapter, the radial fluxes of particles and energy in the $1/\nu-$regime 
are calculated following Ref.~\cite{NemovKasilov:PoP1999} with the mono-energetic 
DKE treated in the relativistic approach.

\Eref{eq:DKEmono} can be solved by integration along the field-line.
Here, only the trapped electrons, $B_0/\Bmax<\lambda<B_0/\Bmin$, are considered 
($\Bmax$ and $\Bmin$ are the absolute maximum and minimum of $B$ on the given 
magnetic surface, respectively).
Enumerating the local minima of $B$ along the magnetic field-line by $k$ and 
integrating \eref{eq:DKEmono} over the bounce trajectory (assumed to be closed), 
one can obtain
\begin{equation}\label{eq:banana}
\dfrac{2\gamma\nu_D(u)}{u}\dfrac{\partial}{\partial\lambda}
\left(\lambda I^{(k)}\dfrac{\partial f_{e1}^{(k)}}{\partial\lambda}\right) = 
\delta\rho^{(k)}\dfrac{\partial F_{eMJ}}{\partial\rho}
\end{equation}
with
\begin{equation}\label{eq:Ik}
I^{(k)}=\oint_{(k)}\dfrac{ds}{b}\xi
\;\;\mathrm{and}\;\;
\delta\rho^{(k)} = \dfrac{\gamma}{u}\oint_{(k)}\dfrac{ds}{\xi}\dot{\rho},
\end{equation}
where $\delta\rho^{(k)}$ is the radial displacement of an electron due to 
the magnetic drift after one bounce period.
To solve \eref{eq:banana}, the following trick was used \cite{NemovKasilov:PoP1999}.
Applying the explicit expression for $\bVdr$ given by \eref{eq:Vdr} to 
$\dot{\rho}=\bVdr\cdot\nabla\rho$ and using the fact that powers of
$\xi=\sigma\sqrt{1-\lambda b}$ with $\sigma = \pm 1$ can be expressed as
\begin{equation}\label{eq:trick}
\xi^m=-\dfrac{2}{(m+2)b}\,\dfrac{\partial}{\partial\lambda}\xi^{m+2},
\end{equation}
one can represent the integrand for $\delta\rho^{(k)}$ in \eref{eq:Ik} 
as follows:
\begin{equation}\label{eq:rhodotxi}
\dfrac{{\dot\rho}}{\xi}=-\dfrac{u^2}{\gamma}\,
\dfrac{|\nabla\rho|k_G}{b^2\omega_{c0}}\,
\dfrac{\partial}{\partial\lambda}
\left(\xi+\dfrac{\xi^3}{3}\right),
\end{equation}
where $\omega_{c0}=eB_0/(m_{e0}c)$ is the cyclotron frequency,
$k_G={\bf n_\rho\cdot[h\times(h\cdot\nabla)h]}$ is the geodesic curvature 
of the magnetic field line and $\bf n_\rho=\nabla\rho/|\nabla\rho|$ is the 
unit vector normal to the magnetic surface. 
Then 
\begin{equation}\label{eq:rho_k}
\delta\rho^{(k)} = -\dfrac{u}{3}\dfrac{\partial H^{(k)}}{\partial\lambda}
\;\;\mathrm{with}\;\;
H^{(k)} = \oint_{(k)} ds\;\xi(3+\xi^2)\dfrac{|\nabla\rho|k_G}{b^2\omega_{c0}}.
\end{equation}
Using this relation and the fact that $I^{(k)}=H^{(k)}=0$ at the bottom 
of the magnetic wells (when $\lambda=B_0/\Bmin$), the order of 
\eref{eq:banana} can be reduced,
\begin{equation}\label{eq:banana2}
\dfrac{\partial f_{e1}^{(k)}}{\partial\lambda} =
-\dfrac{H^{(k)}}{6\lambda I^{(k)}}\,\dfrac{u^2}{\gamma\nu_D(u)}\,
\dfrac{\partial F_{eMJ}}{\partial\rho}.
\end{equation}
The radial components of the particle and energy fluxes are given by 
\begin{subequations}
\label{flux1}
\begin{eqnarray}
\Gamma_e^\rho &=& \langle{\bf \Gamma}_e\cdot\nabla\rho\rangle = 
\left\langle\int d^3u\,\dot{\rho}\,f_{e1}\right\rangle,        \label{Gr}\\
Q_e^\rho &=& \langle{\bf Q}_e\cdot\nabla\rho\rangle = 
\left\langle\int d^3u\,m_{e0}c^2(\gamma-1)\,\dot{\rho}\,f_{e1}\right\rangle, 
\label{Qr}
\end{eqnarray}
\end{subequations}
where $f_{e1}$ is the solution of the relativistic drift-kinetic equation 
\eref{eq:DKEmono} and $\langle ...\rangle$ means averaging on the magnetic surface.

The conductive heat flux for relativistic electrons requires special attention.
According to its physical definition
\cite{HelanderSigmar:CollisTransport_2002,HintonHazeltine:RMP1976}, 
the radial conductive heat flux can be found by extracting the advective and 
mechanical contributions from the radial component of the energy flux, 
\begin{equation}\label{eq:qe_phys}
q_e^\rho=Q_e^\rho-T_e\Gamma_e^\rho-W_e V^\rho, 
\end{equation}
where 
\begin{equation}\label{eq:We}
W_e=\int d^3u\,m_{e0}c^2(\gamma-1)\,f_{eMJ}=\left(\frac{3}{2}+\cR\right)n_eT_e
\end{equation}
is the energy density related to the Maxwell-J\"uttner distribution function, 
and $V^\rho=\Gamma_e^\rho/n_e$ is the flow velocity.
Finally, the radial heat flux can be written as
\begin{equation}\label{eq:qe}
q_e^\rho=Q_e^\rho-\left(\dfrac{5}{2}+\cR\right)T_e\Gamma_e^\rho.
\end{equation}
Please note that this definition differs from the non-relativistic expression 
accepted in the neoclassical theory \cite{HelanderSigmar:CollisTransport_2002,
HintonHazeltine:RMP1976} by the additional correction term $\cR$.

It is convenient to use common notations for both the particle and energy 
fluxes of the form $J_i=\left<\int h_i\dot{\rho}f_{e1}d^3u\right>$,  
where $J_1\equiv\Gamma_e^\rho$ and $J_2\equiv Q_e^\rho/T_e$ 
with $h_1=1$ and $h_2=\kappa\equiv\mu_r(\gamma-1)$, respectively.
Using in $\int d^3u$ the variable $\kappa$ instead of $u$ and performing 
the integration over $\lambda$ instead of pitch, 
$\int d\xi=b/2\sum_\sigma\int d\lambda/|\xi|$, one can obtain from \eref{flux1} 
the following:
\begin{equation}\label{eq:flux2}
J_i = \dfrac{\pi}{2}\ute^3\int_0^\infty d\kappa\sqrt{\kappa}\,h_i
\gamma\left(\dfrac{\gamma+1}{2}\right)^{1/2}
\left<b\sum\limits_{\sigma=\pm 1}
\int_{1/b_\mathrm{max}^{(k)}}^{1/b_\mathrm{min}^{(k)}}d\lambda\,
f_{e1}^{(k)}\dfrac{\dot{\rho}}{|\xi|}\right>.
\end{equation}
Then, substituting \eref{eq:rhodotxi} into \eref{eq:flux2}, an integration 
by parts over $\lambda$ can be performed. 
Finally, considering the averaging over the flux-surface as the limit of 
integration along the field-line and applying \eref{eq:banana2}, the 
desired expression for the radial fluxes can be obtained, 
\begin{equation}\label{eq:flux3}
J_i = -n_e G_0 C_{MJ}\dint_0^\infty d\kappa\,
\dfrac{\e^{-\kappa}\kappa^{5/2}}{\gamma\hat{\nu}_D(u)}
\left(\dfrac{\gamma+1}{2}\right)^{5/2}h_i\,
\dfrac{\partial\ln F_{eMJ}}{\partial\rho},
\end{equation}
with $\hat{\nu}_D(u)\equiv\nu_D(u)/\nu_{e0}$. 
One can check that the non-relativistic limit considered in 
Ref.~\cite{NemovKasilov:PoP1999} is recovered.

The coefficient $G_0$ in \eref{eq:flux3}, identical for both relativistic 
and non-relativistic formulations, contains all parameters for plasmas and 
magnetic configuration which are specific for the considered $1/\nu-$regime,
\begin{equation}\label{eq:geomfactor}
G_0 = \dfrac{4\sqrt{2}}{9\pi^{3/2}}\,
\dfrac{\ute^4}{R^2\omega_{e0}^2\nu_{e0}}\,\left<|\nabla\rho|\right>^2
\epsilon_\mathrm{eff}^{3/2},
\end{equation}
where $R$ is the major radius and $\epsilon_\mathrm{eff}$ is the effective 
ripple amplitude (not shown here; for details see \cite{NemovKasilov:PoP1999}).
The expression for the radial fluxes in the $1/\nu$ collisional regime calculated 
in the relativistic approach \eref{eq:flux3} is, actually, the main result of 
this paper.

Substituting in \eref{eq:flux3} the derivative of the Maxwell-J\"uttner 
distribution function \eref{eq:RHS}, the radial fluxes can be expressed as
\begin{equation}\label{eq:genflux}
J_i=-n_e\sum_{j=1,2} L_{ij}A_j,
\end{equation}
where the thermodynamic forces $A_1$ and $A_2$ are defined in \eref{eq:forcesA1A2}
and the transport coefficients are equal to 
\begin{equation}\label{eq:Lij}
L_{ij}=G_0C_{MJ}\dint_0^\infty d\kappa\,
\dfrac{\e^{-\kappa}\kappa^{5/2}}{\gamma\hat{\nu}_D(u)}
\left(\dfrac{\gamma+1}{2}\right)^{5/2}h_ih_j,
\end{equation}
with $i,j = 1,2$. This definition satisfies Onsager symmetry.

\section{Comparison of relativistic and non-relativistic radial fluxes}
\label{sec:compare}

In this chapter, the role of relativistic effects in the radial neoclassical 
transport is examined. 
Using relativistic expression for $\nu_D(u)$ (see \ref{sec:nuDab}), direct 
numerical integration in \eref{eq:Lij} can be done.
For comparison, the expression for non-relativistic transport is appropriate.
The latter can be obtained from \eref{eq:Lij} by letting $C_{MJ}=\gamma=1$, 
$\kappa=v^2/\vte^2$, and with the non-relativistic expression for $\nu_D(v)$.
Since the geometrical part of the transport coefficients is the same for both 
non-relativistic and relativistic approaches, the ratio of these quantities 
is a pure indicator of the relativistic effects.

\begin{figure}[!tbp]
\begin{center}
\includegraphics[scale=0.97,angle=0]{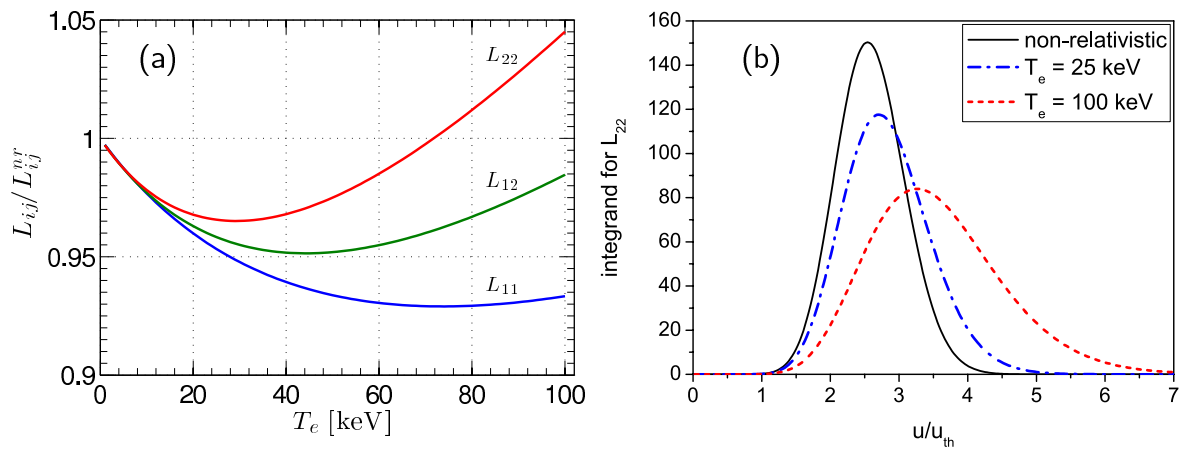}
\caption{[Color online] a) Relativistic transport coefficients $L_{ij}$ divided 
by corresponding non-relativistic values $L_{ij}^{nr}$ are shown as the 
function of electron temperature.
b) Integrand of $L_{22}$ (see \eref{eq:Lij}) is plotted for the different 
temperatures. 
The value $T_e=1$ eV is taken as the non-relativistic limit.}
\label{fig:Lij}
\end{center}
\end{figure}

In \fref{fig:Lij}(a), the ratios $L_{ij}/L_{ij}^\mathrm{nr}$ are shown
as a functions of $T_e$ for $i,j = 1,2$ (here and below, the label 
``$\mathrm{nr}$'' indicates the non-relativistic quantities).
One can see that the correction provided by the relativistic effects is 
not very strong (less than 7\% for this range of temperature).
However, a non-monotonic temperature dependence is not intuitively 
expected and requires an interpretation.

In \fref{fig:Lij}(b), the integrand for $L_{22}$ is plotted as a function 
of $u/u_{te}$ for different temperatures.
One can see that a non-monotonic temperature dependence of 
$L_{ij}/L_{ij}^\mathrm{nr}$ can be explained by superposition of two 
counteracting relativistic effects.
The first one appears due to a reduction of the contribution from the bulk 
of the distribution function and prevails in the low-temperature range, 
$T_e<10$ keV, leading to a decrease of transport coefficients 
(note that the slope in \fref{fig:Lij}(a) is almost the same for all transport 
coefficients in this temperature range). 
A decrease of the bulk contribution is caused by the specific feature of 
the Maxwell-J\"uttner distribution function and can be estimated from 
$C_{MJ}$ \eref{eq:CMJ}.
The second effect is caused by a broadening of the energy-range of 
contributing electrons and the shift of the maximum of the integrand 
into higher energies, and this leads to an increase of the transport 
coefficients with temperature.
The latter effect appears to be important at higher temperatures. 
As one can see from \fref{fig:Lij}(b), in the non-relativistic limit 
($T_e=1$~eV) the major contribution is coming from the electrons with 
$u/\ute\sim 1-4$, while at higher temperatures this range becomes broader 
and for $T_e=100$~keV the corresponding range is $u/\ute\sim 1.5-6.5$.
This effect is weaker for $L_{11}$ than for $L_{12}$ and $L_{22}$ 
due to the lower power of $\kappa$ in the integrand in \eref{eq:Lij}.

Unlike the non-relativistic case, the transport coefficients do not fully 
characterize the transport properties of a confined plasma (because of the 
relativistic factor $\cR$ in $A_1$) and, consequently, the comparison of 
particle and energy fluxes is necessary as well.
In order to make a comparison with the non-relativistic limit possible, 
let us consider two special cases: 
($a$) $n_e^\prime=\Phi^\prime=0$, and ($b$) $T_e^\prime=0$, 
and the corresponding fluxes can be written, respectively, 
as
\begin{subequations}
\label{eq:Jab}
\begin{eqnarray}
J_i^{(a)}&=&-n_e\left[-\left(\dfrac{3}{2}+\cR\right)L_{i,1}+L_{i,2}\right]
\dfrac{T_e^\prime}{T_e}, \label{Ja}\\
J_i^{(b)}&=&-n_e L_{i,1}\left(\dfrac{n_e^\prime}{n_e}-
\dfrac{e\Phi^\prime}{T_e}\right). \label{Jb}
\end{eqnarray}
\end{subequations}
In both cases, the ratio $J_i/J_i^\mathrm{nr}$ no longer contains the 
gradients and can be easily calculated.
Note that in the case ($b$), 
$\Gamma_e^\rho/\Gamma_e^{\rho,\mathrm{nr}} = L_{11}/L_{11}^{\mathrm{nr}}$,
i.e. the relativistic correction for $\Gamma_e^\rho$ is identical to $L_{11}$ 
which is shown in \fref{fig:Lij}(a).

Following Eqs.~(\ref{eq:qe}) and (\ref{eq:Jab}), 
the heat flux can also be represented in a similar manner:
\begin{subequations}
\label{eq:qab}
\begin{eqnarray}
q_e^{\rho,(a)}/T_e &=& -n_e\left[-\left(4+2\cR\right)L_{12}+L_{22}
 +\left(\dfrac{15}{4}+4\cR+\cR^2\right)L_{11}\right]
\dfrac{T_e^\prime}{T_e}, \label{qa}\\
q_e^{\rho,(b)}/T_e &=& -n_e\left[L_{12}-\left(\dfrac{5}{2}+\cR\right)L_{11}\right]
\left(\dfrac{n_e^\prime}{n_e}-\dfrac{e\Phi^\prime}{T_e}\right).\label{qb}
\end{eqnarray}
\end{subequations}
In \eref{eq:qab}, the Onsager symmetry, $L_{12}=L_{21}$, was used.

\begin{figure}[!tbp]
\begin{center}
\includegraphics[scale=0.78,angle=0]{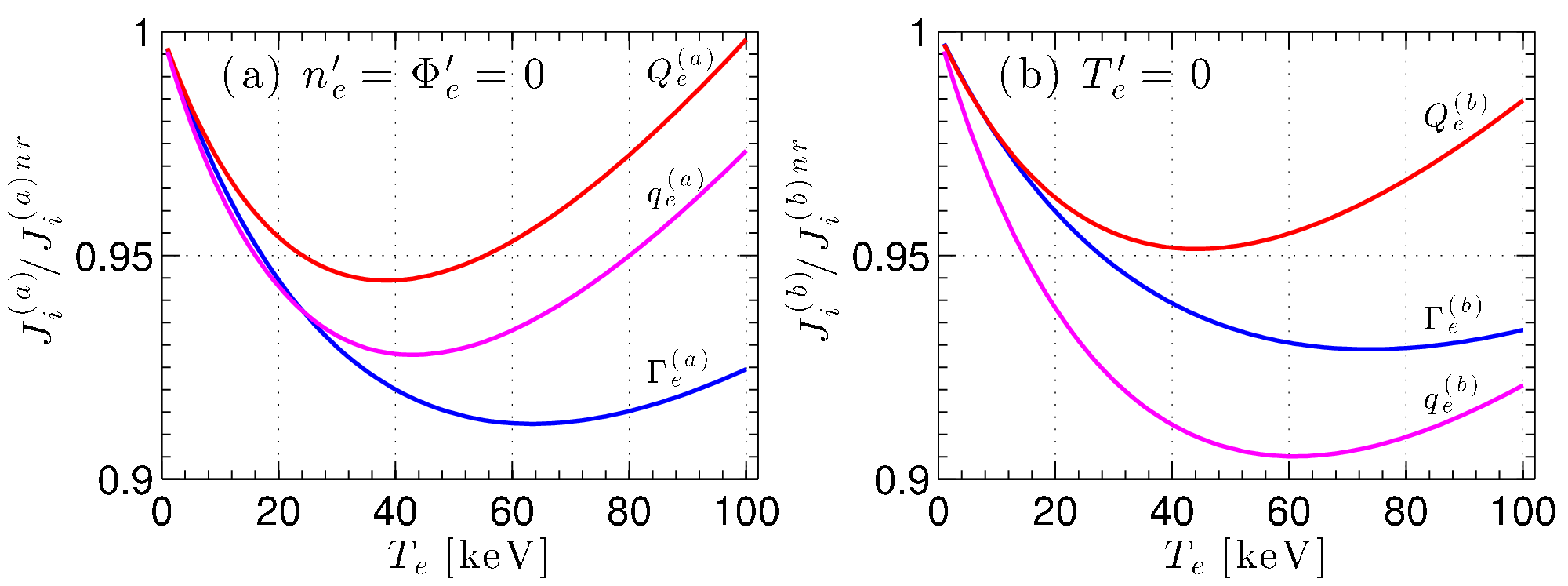}
\caption{[Color online] 
The temperature dependence of the ratios $J_i/J_i^{nr}$ for two cases, 
(a) and (b), respectively, is shown.}
\label{fig:fluxes}
\end{center}
\end{figure}

In \fref{fig:fluxes}, the ratios of $\Gamma_e^\rho/\Gamma_e^{\rho,\mathrm{nr}}$,
$Q_e^\rho/Q_e^{\rho,\mathrm{nr}}$ and $q_e^\rho/q_e^{\rho,\mathrm{nr}}$ 
for both cases are shown.
The same non-monotonic dependence as in the case of the transport coefficients 
is clearly indicated, and relativistic correction for the $1/\nu$ radial fluxes 
is found to be less then $10\%$ for the temperature range checked.

\section{Summary}
\label{sec:sum}

In this paper, the neoclassical radial fluxes for hot electrons in the $1/\nu$ 
regime, which is specific to stellarators, has been calculated in the relativistic 
approach. 
The choice of $1/\nu$ regime was motivated by a rather intuitive expectation 
that the role of relativistic effects in this regime should be most pronounced 
since the contribution to the radial transport from the tail of the distribution 
function is the largest (the integrand in the transport coefficients for the 
non-relativistic electrons in this regime scales in lowest order as $\propto v^7$).

For calculations, the reduced mono-energetic relativistic DKE was derived.
Apart from the radial particle and energy fluxes, also the expression for 
the relativistic conductive heat flux was obtained.
The definition for the radial fluxes in the relativistic approach has an 
important feature: 
the relativistic effects enter in the fluxes not only through the distribution 
function, but also through an additional temperature-dependent term 
in the first thermodynamic force.
This relativistic term depends only on the temperature, in contrast to the 
canonical set of radial thermodynamic forces in which the logarithmic gradients 
of plasma parameters appear. 
Nevertheless, use of the proposed formulation has a big advantage: 
the transport coefficients with the relativistic effects taken into account 
can be calculated by the same numerical solvers which solve the non-relativistic 
DKE directly.

Following Ref.~\cite{NemovKasilov:PoP1999}, the radial fluxes were calculated 
from the relativistic mono-energetic DKE and the results obtained were compared 
with the corresponding non-relativistic quantities.
It was found that the relativistic effects for hot electrons produce a modest, 
but systematic reduction of the radial transport (up to 10\% within the temperature 
range relevant for fusion).
However, a non-monotonic temperature dependence of the transport coefficients 
is somewhat surprising. 
This behavior is the result of two counteracting factors present for 
relativistic kinetics.
The first factor is related to a reduction in the relativistic Maxwellians of 
the weight of bulk electrons with an increase of the temperature.
The second factor is caused by a broadening of the energy-range and a shift of 
the maximum contribution to higher energies.

This initial investigation confirms the intuitive expectation of an absence of 
strong relativistic effects in the radial transport in stellarator fusion plasmas. 
At the same time, this conclusion is not general and a similar check must also 
be made for the banana-regime in tokamaks.
Apart from this, maybe the most important task is the calculation of the parallel 
electron fluxes with the relativistic effects taken into account.
Based on the results provided in this paper, one may expect that within the 
non-relativistic neoclassical treatment both the electron radial fluxes in the 
banana regime and the electron bootstrap current in hot plasmas are 
somewhat overestimated.

\begin{ack} 
The authors would like to acknowledge C.~D.~Beidler and S.~V.~Kasilov for 
support and fruitful discussions and the latter, in particular, for the idea 
to examine the radial fluxes in the $1/\nu$-regime using the developed relativistic 
approach.
\end{ack}
\appendix
\section{Relativistic expressions for deflection frequencies}
\label{sec:nuDab}

Expressing the deflection frequency for the test-particle $a$ immersed in 
the background $b$ through the diffusion coefficient for pitch-angle scattering,
$\nu_D^{ab}(u)=(2/u^2)D^{ab}_{\theta\theta}(u)$, and taking the general 
relativistic definition for $D^{ab}_{\theta\theta}(u)$ from 
Ref.~\cite{BraamsKarney:PoFB1989},
the general expression for $\nu_D^{ee}(u)$ with the relativistic 
Maxwellian can be written as follows:
\begin{equation}\label{eq:nuee}
\begin{array}{lll}
\fl \nu_D^{ee}(u) = \nu_{e0}C_{MJ}(\mu_r)\dfrac{4}{\sqrt{\pi}}\times\\
\fl \qquad\left(\dfrac{\gamma}{u^3}\dint\limits_0^u
\left[\gamma^\prime-2\left(\dfrac{c^2}{u^2}+
\dfrac{1}{\gamma^2}\right)j_{0[2]02}^\prime+
\dfrac{8}{\gamma^2}\dfrac{c^2}{u^2}j_{0[3]022}^\prime\right]
\dfrac{{u^\prime}^2}{\gamma^\prime}\e^{-\mu_r(\gamma^\prime-1)}du^\prime+\right.
\nonumber\\
\fl \qquad\quad\left.\dfrac{\gamma}{u^2}\dint\limits_u^\infty
\left[\dfrac{{\gamma^\prime}^2}{\gamma}
-2\left(\dfrac{c^2}{u^2}+
\dfrac{{u^\prime}^2}{u^2\gamma^2}\right)j_{0[2]02}+
\dfrac{8}{\gamma^2}\dfrac{c^2}{u^2}j_{0[3]022}\right]
\dfrac{u^\prime}{\gamma^\prime}\e^{-\mu_r(\gamma^\prime-1)}du^\prime\right).
\end{array}
\end{equation}
The specific functions $j_{l[k]*}(z)$ \cite{BraamsKarney:PoFB1989}
are given by:
\begin{equation}\label{eq:j_xx}
\begin{array}{lll}
j_{0[2]02}(z) &=& (z\gamma-\sigma)/4z,\\
j_{0[3]022}(z) &=& [-3z\gamma+(3+2z^2)\sigma]/32z,
\end{array}
\end{equation}
where $\sigma(z)=\ln(z+\gamma)$ with $\gamma=\sqrt{1+z^2}$ and $z=u/c$.
Since the leading order for $z\ll1$ is
$j_{0[2]02}\simeq z^2/6$ and $j_{0[3]022}\simeq z^4/120$,
the non-relativistic limit, $c\rightarrow\infty$, can be easily obtained 
\cite{HintonHazeltine:RMP1976,HelanderSigmar:CollisTransport_2002},
\begin{equation}\label{eq:nuee_nr}
\nu_D^{ee}(v)=\nu_{e0}\dfrac{1}{x^3}
\left[\left(1-\dfrac{1}{2x^2}\right)\mathrm{erf}(x)+
\dfrac{\mathrm{erf}^\prime(x)}{2x}\right],
\end{equation}
where $x=v/\vte$.

For $\nu_D^{ei}$, the ion background can be taken as a non-relativistic Maxwellian.
For calculations, it is sufficient to apply the high-speed-limit:
\begin{equation}\label{eq:nuei}
\nu_D^{ei}(u)=\nu_{e0}\Zeff\,\dfrac{\gamma\ute^3}{u^3}.
\end{equation}
One can see that the non-relativistic limit for this expression also exactly
coincides with the classical approach.

\section*{References}
\bibliographystyle{prsty}
\bibliography{../../../../biblio/biblio}
\end{document}